\documentclass[reprint, superscriptaddress, aps, 10pt]{revtex4-2}

\usepackage[ngerman, english]{babel}
\usepackage[T1]{fontenc} 

\usepackage{amsmath, amssymb, amsthm}
\usepackage{tensor}
\usepackage{braket}
\usepackage{bm}
\usepackage{IEEEtrantools}
\usepackage{siunitx}
\DeclareSIUnit\year{yr}

\usepackage{placeins}
\usepackage{booktabs}
\newcolumntype{L}[1]{>{\raggedright\let\newline\\\arraybackslash\hspace{0pt}}p{#1}}
\newcolumntype{C}[1]{>{\centering\let\newline\\\arraybackslash\hspace{0pt}}p{#1}}
\newcolumntype{R}[1]{>{\raggedleft\let\newline\\\arraybackslash\hspace{0pt}}p{#1}}
\usepackage[normalem]{ulem}
\usepackage[svgnames]{xcolor}
\usepackage{hyperref}
\hypersetup{
    colorlinks=true,
    citecolor=NavyBlue,
    filecolor=black,
    linkcolor=NavyBlue,
    urlcolor=NavyBlue,
    linktocpage=true
}
\usepackage{pgf}
\usepackage{graphicx}

\usepackage[capitalize]{cleveref}
\Crefname{figure}{Fig.}{Figs.}
\Crefname{equation}{Eq.}{Eqs.}

\usepackage[acronym,symbols,nogroupskip,nomain,nonumberlist,nopostdot,toc]{glossaries}
\glsdisablehyper
\newacronym{vp}{VP}{variational principle}
\newacronym{vqa}{VQA}{variational quantum algorithm}
\newacronym{vqe}{VQE}{variational quantum eigensolver}
\newacronym{tdvp}{TDVP}{time-dependent variational principle}
\newacronym{mvp}{MVP}{McLachlan variational principle}
\newacronym{dfvp}{DFVP}{Dirac-Frenkel variational principle}
\newacronym[firstplural=degrees of freedom]{dof}{DOF}{degree of freedom}
\newacronym{ode}{ODE}{ordinary differential equation}
\newacronym{tdse}{TDSE}{time-dependent Schrödinger equation}
\newacronym[firstplural=equations of motion (EOMs)]{eom}{EOM}{equation of motion}
\newacronym{onv}{ONV}{occupation number vector}
\newacronym{mctdh}{MCTDH}{multiconfigurational time-dependent Hartree}
\newacronym{vha}{VHA}{Variational Hamiltonian Ansatz}
\newacronym{pt}{PT}{polaron transformation}

\def\bea{\begin{IEEEeqnarray}}
\def\eea{\end{IEEEeqnarray}}
\newcommand{\diff}{\mathrm{d}}
\newcommand{\order}{\mathcal{O}}

\def\a{a{\mathstrut}}
\def\ad{a{\mathstrut}^\dagger}
\def\ta{\tilde{a}{\mathstrut}}
\def\tad{\tilde{a}{\mathstrut}^\dagger}

\def\si{\sigma{\mathstrut}^i}
\def\sx{\sigma{\mathstrut}^x}
\def\sy{\sigma{\mathstrut}^y}
\def\sz{\sigma{\mathstrut}^z}
\def\sm{\sigma{\mathstrut}^-}
\def\sp{\sigma{\mathstrut}^+}
\def\nmax{n{\mathstrut}^\mathrm{max}}

\begin{document}

\title{Quantum algorithms for quantum dynamics: A performance study on the spin-boson model}

\author{Alexander Miessen}
\affiliation{IBM Quantum, IBM Research -- Z{\"u}rich, 8803 R{\"u}schlikon, Switzerland}
\affiliation{Institute for Theoretical Physics, ETH Z{\"u}rich, 8093 Z{\"u}rich, Switzerland}

\author{Pauline J. Ollitrault}
\affiliation{IBM Quantum, IBM Research -- Z{\"u}rich, 8803 R{\"u}schlikon, Switzerland}

\author{Ivano Tavernelli}
\affiliation{IBM Quantum, IBM Research -- Z{\"u}rich, 8803 R{\"u}schlikon, Switzerland}

\date{\today}

\begin{abstract}
Quantum algorithms for quantum dynamics simulations are traditionally based on implementing a Trotter-approximation of the time-evolution operator.
This approach typically relies on deep circuits and is therefore hampered by the substantial limitations of available noisy and near-term quantum hardware.
On the other hand, \glspl{vqa} have become an indispensable alternative, enabling small-scale simulations on present-day hardware.
However, despite the recent development of \glspl{vqa} for quantum dynamics, a detailed assessment of their efficiency and scalability is yet to be presented.
To fill this gap, we applied a \gls{vqa} based on McLachlan's principle to simulate the dynamics of a spin-boson model 
subject to varying levels of realistic hardware noise as well as in different physical regimes, and discuss the algorithm's accuracy and scaling behavior as a function of system size.
We observe a good performance of the variational approach used in combination with a general, physically motivated wavefunction ansatz, and compare it to the conventional first-order Trotter-evolution.
Finally, based on this, we make scaling predictions for the simulation of a classically intractable system.
We show that, despite providing a clear reduction of quantum gate cost, the variational method in its current implementation is unlikely to lead to a quantum advantage for the solution of time-dependent problems.
\end{abstract}

\maketitle

\newpage

\section{Introduction}
\label{sec:intro}

The simulation of quantum systems is one of the most promising applications of quantum computing \cite{Zalka1998}, aiming to overcome the limits of classical computers when it comes to storing and manipulating exponentially large quantum states.
However, many of the conceived quantum algorithms, claiming to offer exponential speed-up over classical counterparts, are too resource-intensive for available hardware and will only become practicable once fault-tolerance is reached.
In turn, since today's noisy near-term quantum technology is characterised by low qubit counts ($<1000$), short decoherence times ($\sim \SI{100}{\micro\second}$) and two-qubit gate errors ($\sim 10^{-3}$) \cite{ibmqx, Preskill2018}, error-correction schemes cannot yet be implemented~\cite{aharonov2008fault}.

This has sparked the development of hybrid quantum-classical algorithms, or \glspl{vqa}~\cite{Bharti2021}, that split the workload between a quantum and a classical processor.
Most prominently, the \gls{vqe} has become the standard-tool for eigenvalue problems~\cite{Peruzzo2014, McClean2016, Moll2018}.
With efficient encodings of variational states, \gls{vqe} requires only shallow circuits and has enabled small-scale simulations of up to a few atoms already on present-day hardware~\cite{Kandala2017, Kandala2019, Ollitrault2020b}.

Since the development of a first \gls{vqa} for quantum dynamics by Li \textit{et al.} in 2017~\cite{Li2017}, there has been a surge in attention to the simulation of quantum dynamics using \glspl{vqa}.
Several new methods, partially based on Ref.~\citenum{Li2017}, have been put forward recently~\cite{Heya2019, Cirstoiu2020, Zhang2020, Yao2020, Bharti2020, Barison2021, Lau2021, Lau2021a, Benedetti2020}.
These approaches claim to be more resource-efficient compared with fault-tolerant quantum algorithms for implementing the time evolution operator, $U_t = e^{-iHt}$, such as product formulas for the decomposition of $U_t$, commonly known as Trotter formulas~\cite{Lloyd1996, Kassal2008, Smith2019, Chiesa2019, Ollitrault2020a}, linear combination of unitaries~\cite{Childs2012hamiltonian}, quantum signal processing \cite{Low2017optimal}, and qubitization \cite{Low2019hamiltonian}.

However, for \glspl{vqa} to be meaningful for near-term applications in the simulation of quantum dynamics, it is necessary to carefully evaluate their performance, including their stability under noisy hardware conditions.
Furthermore, their versatility with different systems has to be assessed.
Particularly, they rely on choosing a variational ansatz that is both compact and flexible enough to accurately represent the studied system during the entire dynamics.
Finding such a variational form is itself highly non-trivial as already addressed in the literature~\cite{Zhang2020}, which is why often, a so-called heuristic, or hardware-efficient ansatz, is chosen.
Such an ansatz is agnostic to the problem at hand and its underlying symmetries, resulting in high numbers of variational parameters which could potentially jeopardize desired quantum advantage.
Hence, in order to better characterize these \glspl{vqa}, their application to non-trivial systems~\cite{Lee2021} is essential.

In this work we propose a detailed study of the performance of Li's \gls{vqa}~\cite{Li2017} for solving the dynamics of a spin-boson model.
Moreover, based on our results, we make predictions on scalability and possible quantum advantage with a particular focus on the comparison with Trotter-evolution.
The spin-boson model presents itself as an ideal testbed due to its rich dynamics and high relevance for various areas of research, resulting in a multitude of theoretical~\cite{Yao2013, Peropadre2013, Diaz-Camacho2016} and experimental studies~\cite{Mezzacapo2014, Braumuller2017, Langford2017}.
The generic model of a two-level system coupled to a bath of harmonic oscillators is of great importance in the study of light-matter interaction and, particularly so, in the description of optical cavities and superconducting circuits~\cite{FriskKockum2019}.
On the other hand, it may also be seen as an idealized model for the study of the non-adiabatic dynamics of molecules, where, in this case, the fermionic two-level system describes two molecular potential energy surfaces~\cite{Ollitrault2020a, Tong2020}.
Recent efforts in the context of digital quantum computing have explored both the spin-boson model's stationary as well as dynamical properties~\cite{Macridin2018, DiPaolo2020, Fitzpatrick2021}.

In this work, we start by constructing a physically motivated time-dependent variational form.
We then focus on the numerical stability of the algorithm in different physical regimes, as well as the effects of introducing realistic experimental noise.
In the last section, we finally present a careful study on the scaling of the computational resources as a function of the system size, comparing the variational approach and Trotter-evolution.
In particular, we present predictions for system sizes far out of reach for classical simulation and conclude on the possibility to reach quantum advantage using near-term and fault-tolerant quantum algorithms for quantum dynamics.

\section{Theory}
\label{sec:theory}

\subsection{Quantum dynamics with product formulas}
\label{subsec:productformulas}

As eluded to in the introduction, the most widely used method for time-evolution in the context of quantum computing remains the approximation of the unitary time evolution operator with a Trotter-Suzuki formula.
At first order and with $H = \sum_{j=1}^{N_\mathrm{h}} h_j$, we have
\begin{equation}
    \exp(-iHt) \approx \Bigl( \prod_{j=1}^{N_\mathrm{h}} e^{-i h_j \frac{t}{d}} \Bigr)^d \ ,
\label{eq:trotter}
\end{equation}
with an error that scales with $\order(N_\mathrm{h}^2t^2 / d)$.
It can be shown, however, that for Hamiltonians which can be mapped to a qubit-lattice and split into even and odd parts, as is the case for the spin-boson Hamiltonian introduced below, this scaling reduces to linear in the number of Hamiltonian terms \cite{Chiesa2019, Childs2019a},
\begin{equation}
    \varepsilon = \order \Bigl( N_\mathrm{h} \frac{t^2}{d} \Bigr) \ .
\label{eq:trotterScaling}
\end{equation}
The drawback of this method is that it typically requires long circuits due to the error scaling quadratically with the simulation time.

\subsection{Variational quantum algorithm for real time evolution}
\label{subsec:algorithm}

Alternatively, variational time-evolution algorithms for quantum dynamics aim to drastically reduce the circuit depth.
A time-dependent variational ansatz $\ket{\Phi(\bm{\theta})}$, with $\bm{\theta} = \bm{\theta}(t)$, seeks to approximate the true state $\ket{\Psi(t)}$, obtained as a solution to the \gls{tdse} $i \hbar \frac{\diff \ket{\Psi}}{\diff t} = H \ket{\Psi}$. The parameter's time-dependence will be left implicit in the following and we set $\hbar = 1$.

On a quantum computer, a variational ansatz is prepared by acting upon a reference qubit-state $\ket{\phi}$ with a parameterized unitary operator, the quantum circuit, $\ket{\Phi(\bm{\theta})} = U(\bm{\theta}) \ket{\phi}$, where $\bm{\theta} = (\theta_1, \theta_2, \ldots) \in \mathbb{R}^{N_\theta}$ is a set of real parameters.
Although variational parameters can generally be complex, they are, in fact, required to be real in the setting of quantum computation since they will be encoded as angles of rotational quantum gates.
As outlined in \cite{Li2017, Yuan2019}, such a time-dependent varational ansatz can be employed in a hybrid quantum-classical algorithm.

One of three \glspl{vp} \cite{Hackl2020, Martinazzo2020, Broeckhove1988}, the \gls{dfvp}~\cite{Dirac1930, Frenkel1934}, the \gls{mvp}~\cite{McLachlan1964}, and the \gls{tdvp}~\cite{Kramer1981}, may then be used to derive a set of \glspl{eom} dictating the parameter evolution.
In fact, as is intelligibly shown in \cite{Broeckhove1988}, all three principles are equivalent under the condition that the variational manifold $\mathcal{M}$ is such that $\ket{\delta \Phi}$ and $i \ket{\delta \Phi}$ are both elements of the same tangent space.
This is typically satisfied for a complex parameterization but not for purely real parameters \cite{Hackl2020}, as is the case here.
In fact, while parameters have to be made real artificially with the \gls{dfvp}, both the \gls{mvp} and the \gls{tdvp} naturally maintain a real parameterization~\cite{Yuan2019, Hackl2020}.
Due to known instabilities in the integration of the \glspl{eom} resulting from the \gls{tdvp}, we will make use of \gls{mvp},
\begin{equation}
    \delta \lVert i \ket{\Theta} - H \ket{\Phi} \rVert = 0 \ ,
\label{eq:McLachlanPrinciple}
\end{equation}
where variation is with respect to $\ket{\Theta} = \ket{\dot{\Phi}}$.
Assuming the evolution of $\ket{\Phi}$ to be governed by the same \gls{tdse} as that of $\ket{\Psi}$, this means to minimize the distance between the projection $H \ket{\Phi}$ and the variational tangent vector $\diff \ket{\Phi} / \diff t$.
\Cref{eq:McLachlanPrinciple} results in the condition $\Im \braket{\delta \Phi | i \partial_t - H | \Phi} = 0$.

With all time-dependence residing in the parameters $\bm{\theta}$ and accounting for a potential global phase mismatch between exact and approximate state, i.e. taking $\ket{\Phi} \rightarrow e^{i \alpha(t)} \ket{\Phi}$, we obtain as \glspl{eom}~\cite{Li2017,Yuan2019, Hackl2020}
\begin{equation}
    \mathcal{M} \dot{\bm{\theta}} = \bm{\mathcal{V}} \ ,
\label{eq:McLachlanEOM}
\end{equation}
with the matrix elements
\begin{equation}
    \mathcal{M}_{ij} = \Re \biggl( \frac{\partial \bra{\Phi}}{\partial \theta_i} \frac{\partial \ket{\Phi}}{\partial \theta_j} + \frac{\partial \bra{\Phi}}{\partial \theta_i} \ket{\Phi} \frac{\partial \bra{\Phi}}{\partial \theta_j} \ket{\Phi} \biggr)
\end{equation}
and the vector components
\begin{equation}
    \mathcal{V}_i = \Im \biggl( \frac{\partial \bra{\Phi}}{\partial \theta_i} H \ket{\Phi} - \frac{\partial \bra{\Phi}}{\partial \theta_i} \ket{\Phi} \braket{\Phi | H | \Phi} \biggr) \ ,
\end{equation}
where the respective second term results from the inclusion of a global phase.
\Cref{eq:McLachlanEOM} may then be solved by any numerical ODE-solver, e.g., a Runge-Kutta method.

We highlight that the \gls{mvp} and the previous derivation is not immanent to quantum computation but may be used for any classical variational ansatz.
What is distinct in the quantum setting is the preparation of the ansatz and the evaluation of individual terms by means of quantum circuits \cite{Somma2002, Schuld2019, Li2017}.

\textcolor{black}{Recently, several other \glspl{vqa} for quantum dynamics were proposed, relying either on propagating parameters by means of an \gls{eom} like \Cref{eq:McLachlanEOM} but differing in the way the ansatz is constructed~\cite{Yao2020, Bharti2020, Lau2021}, or by carrying out an optimization at each timestep~\cite{Barison2021, Lau2021a, Zhang2020, Heya2019, Cirstoiu2020}.
In this last case, one can minimize for instance the distance between a variational state and the outcome of a small Trotter-step, avoiding the measurement-intensive construction of the matrix elements required in \Cref{eq:McLachlanEOM} as well as its inversion, which is a potential source of numerical instabilities.
Concerning the optimization of variational quantum circuits, although it was shown in Ref.~\citenum{Bittel2021} that such optimization is in general NP-hard due to unresolvable local minima, approximate solutions suffice and can be found efficiently in practical simulations (cf. Solovay-Kitaev theorem~\cite{dawson2005solovay}).
Herein, we will make use of the original variational approach in \Cref{eq:McLachlanEOM}, which solely relies on the integration of an \gls{eom} and does not involve any parameter optimization.}

\begin{figure}[t]
  \centering
  \includegraphics{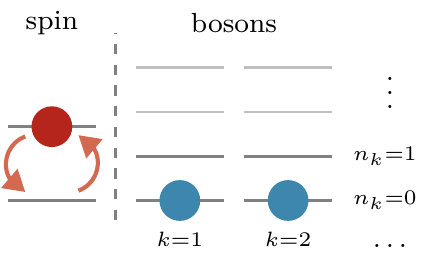}
  \caption{
    Schematic representation of the qubit-mapped spin-boson model.
    The spin's state is captured by a single qubit, while under the direct mapping for bosonic modes, each energy level $n_k$ corresponds to a qubit.
  }
  \label{fig:SBschamic}
\end{figure}

\subsection{The spin-boson model}
\label{subsec:SBmodel}

We consider a two-level system coupled to a bath of $M$ bosons.
The two-level system may represent an atom with two energy levels, a spin-$\frac{1}{2}$ particle, or any artificial system such as, for instance, a superconducting qubit.
For brevity, we will refer to it simply as `the spin'.
Such a system is described by the spin-boson Hamiltonian \cite{FriskKockum2019, DiPaolo2020},
\begin{equation}
    H = \sum_{k=1}^M \omega{\mathstrut}_k \ad_k \a_k
      + \frac{\epsilon}{2} \sz
      + \Delta \sx
      + \sum_{k=1}^M g{\mathstrut}_k \sx (\ad_k + \a_k) \ .
\label{eq:hamiltonian}
\end{equation}
The bosonic operators $\ad_k$ ($\a_k$) create (annihilate) harmonic basis states with eigenfrequencies $\omega_k$, Pauli matrices $\si$, $i \in \{x,y,z\}$ act on the state of the spin with eigenfrequency $\epsilon$ and tunneling rate $\Delta$.
The coupling between spin and bosons is via $\sx$ with coupling constants $g_k$.

Simulation on a quantum device requires to encode states in qubit registers and map operators to quantum gates, e.g., to strings of Pauli operators.
Note that, in the following, the notation will be largely adapted from~\cite{DiPaolo2020}.
The excitation space of the $k$-th bosonic state will be truncated at a maximum occupation number $\nmax_k$, leaving $\nmax_k + 1$ possible occupations per mode $k$, including the ground state.
Under the direct qubit-mapping \cite{Somma2003}, the \gls{onv} is then mapped to a qubit-register of size $\nmax_k + 1$, $\ket{n_k} \longrightarrow \ket{\tilde{n}_k} = \ket{0_{\nmax} \ldots 0_{n_k+1} 1_{n_k} 0_{n_k-1} \ldots 0_{0_k}}$.
Requiring to maintain correct spin-statistics, the corresponing mapping of bosonic creation and annihilation operators follows immediately as $\ad_k \rightarrow \tad_k = \sum_{n_k=0}^{\nmax_k-1} \sqrt{n_k+1} \sp_{n_k} \sm_{n_k+1}$, and analogously for $\a_k$, where $\sigma{\mathstrut}_{n_k}^\pm = (\sx_{n_k} \pm i \sy_{n_k})/2$.

\subsection{The variational ansatz}
\label{subsec:varform}

The so-called \gls{pt} enjoys popularity in the classical simulation of spin-boson models \cite{Diaz-Camacho2016, Shi2018} and has successfully been used for \gls{vqe} ground state calculations recently \cite{DiPaolo2020}.
However, it proved to be insufficient for the use with variational time evolution and the Hamiltonian \Cref{eq:hamiltonian}.
Instead of the \gls{pt}, here we employ a \gls{vha}~\cite{Wecker2015}.
Inspired by the unitary time evolution operator, the time-parameter is simply replaced with a variational parameter that is distinct for each term in $H$, yielding
\bea{rCl}
U_\mathrm{H}(\bm{\theta})
   = \exp \Bigl( - i \Bigl[ & \sum_{k=1}^M\theta_k^{(1)} \ad_k \a_k
       + \theta^{(2)} \sz + \theta^{(3)} \sx \nonumber \\
       & + \sx \sum_{k=1}^M \theta_k^{(4)} (\a_k + \ad_k) \Bigr] \Bigr) \ .
\label{eq:ansatz}
\eea
Note that all Hamiltonian parameters are absorbed into variational parameters.

Translating $U_\mathrm{H}$ into a sum of Pauli strings is now straight-forward.
Employing the above operator mapping, we find
\bea{rCl}
\tad_k \ta_k = \frac{1}{4} \sum_{n_k=0}^{\nmax_k-1} (n_k+1) (& &  \sz_{n_k} - \sz_{n_k+1} - \sz_{n_k} \sz_{n_k+1} \nonumber \\
& & + \frac{1}{2} \mathcal{I}_{n_k} + \frac{1}{2} \mathcal{I}_{n_k+1} ) \ ,
\eea
with $\mathcal{I}$ the identity.
Such identity terms contribute nothing but a global phase upon exponentiation and can thus be neglected in the variational ansatz.

Similarly, for the interaction term, one obtains
\begin{equation}
    \ta_k + \tad_k = \frac{1}{2} \sum_{n_k=0}^{\nmax_k-1} \sqrt{n_k + 1} (\sx_{n_k} \sx_{n_k+1} + \sy_{n_k} \sy_{n_k+1}) \ .
\end{equation}
Since this expression consists of mutually non-commuting terms, the summation over $n_k$ is split into even and odd parts, $X_k^\mathrm{e} := - i \sum_{n_k \mathrm{even}} \sqrt{n_k+1} (\sx_{n_k} \sx_{n_k+1} + \sy_{n_k} \sy_{n_k+1}) / 2$, and analogously for the odd part $X_k^\mathrm{o}$, such that
\begin{equation}
    -i (\ta_k + \tad_k) = X_k^\mathrm{e} + X_k^\mathrm{o} \ .
\label{eq:ansatzCouplingEvenOdd}
\end{equation}
Notably, we have $[X_k^\mathrm{e}, X_k^\mathrm{o}] \neq 0$ while all terms within $X_k^\mathrm{e,o}$ commute.

The resulting exponential is approximated with a Trotter-series of depth $d$, yielding an ansatz suitable for implementation in terms of quantum gates.
Exponentials of Pauli terms may directly be written as rotational gates thereafter, e.g., $R_z (2\theta^{(3)}) = \exp (-i \theta^{(3)} \sz)$.
Although the final variational ansatz appears bulky, it can be compactly expressed as a series of one- and two-qubit gates and may be looked up in \Cref{app:varcirc}.

\subsection{Resource estimates and scaling}
\label{subsec:resources}

In this section, we discuss the scaling of the different computational resources for computing the dynamics of the spin-boson model with both the variational and the Trotter approach, \Cref{eq:trotter} and \Cref{eq:McLachlanEOM}, respectively.
First off, the classical cost per timestep of the variational algorithm is determined by the number of variational parameters, which, for our ansatz (\Cref{eq:ansatz}), is given by
\begin{equation}
    N_\theta = 2d (M n^\mathrm{max} + 1) \ .
\end{equation}

The quantum cost is determined by qubit- and gate-counts, as well as the number of circuit evaluations.
In the variational case, the total number of qubits is
\begin{equation}
    N_\mathrm{q} = M (n^\mathrm{max} + 1) + 1 + 1 \ ,
\end{equation}
where an extra qubit was added to account for the possibility of evaluating gradients by means of an ancilla qubit \cite{Schuld2019}.
Trotter evolution does not require any ancilla, hence requiring one qubit less, $N_\mathrm{q} - 1$.

The number of CNOT gates in the quantum circuit is ansatz-dependent and, for \Cref{eq:ansatz}, can be estimated as
\begin{equation}
    N_\mathrm{cx} = \order \bigl( dMn^\mathrm{max} N_\mathrm{q}\bigr) = \order \bigl( d[Mn^\mathrm{max}]^2 \bigr) \ .
\end{equation}
Assuming the worst qubit-connectivity, i.e., a linear chain, we included a factor $N_\mathrm{q}$ to account for swap gates that enter the circuit upon transpilation.
This means, in the worst-case scenario, one needs to swap over the entire qubit register to execute a CNOT gate.
This is true for both variational and Trotter simulation and, since our ansatz and the Trotter circuit differ only in the gate angles (which are variational parameters in the case of variational simulation), $N_\mathrm{cx}$ is the same for both Trotter and variational simulation.
Note, however, that the Trotter depth, $d$, differs in the two approaches; In the case of the Trotter algorithm, the circuit depth increases quadratically with the simulation time (cf. \Cref{eq:trotterScaling}), while for the variational approach, the depth (and the corresponding number of variational parameters) determines the size and nature of the sub-manifold governing the dynamics.

The number of circuit evaluations per timesteps to evaluate the elements of $\mathcal{M}$ and $\mathcal{V}$ in \Cref{eq:McLachlanEOM} is determined by the number of Hamiltonian terms $N_\mathrm{h}$, the number of circuits necessary to evaluate all gradients $\partial_i \ket{\Phi}$, which we denote $N_{\diff \theta}$, and the number of samples per circuit, $N_\mathrm{shots} = \order(1/\varepsilon^2)$,
\begin{equation}
    N_\mathrm{circ} = \order \Bigl( N_\mathrm{shots} \bigl( N_{\diff \theta}^2 + N_\mathrm{h} N_{\diff \theta} \bigr) \Bigr) \ .
\end{equation}
In the variational ansatz \Cref{eq:ansatz}, we have $N_\theta = 2d (M\nmax + 1)$ variational parameters, a total of $N_{\diff \theta} = d(5M\nmax + 2)$ gradient circuits, and $N_\mathrm{h} = 7M\nmax+2$ Hamiltonian terms.
The number of gradient terms differs from the number of parameters, as some parameters are repeated in the circuit.

Finally, we estimate the number of timesteps taken by the ODE solver to reach the final time $T$.
Throughout this work, we will use an adaptive solver, however, adaptively choosing a step size is highly system dependent.
Therefore, to simplify the estimate, we base it on the local error of a non-adaptive Runge-Kutta solver.
For an order-$p$ Runge-Kutta solver and a fixed timestep $\tau$, the local error scales as $\varepsilon_\mathrm{local}= \order(\tau^{p+1})$.
For a desired final accuray of $\varepsilon_T$, we thus estimate
\begin{equation}
    N_t = \frac{\varepsilon_T}{\varepsilon_\mathrm{local}}
\label{eq:timesteps}
\end{equation}
timesteps.
With $N_t = T/\tau$, this means the timestep must satisfy $\tau^p = \order (\varepsilon_T / T)$.
We emphasize that this is indeed a very rough estimate as scaling coefficients of the local error may heavily depend on the system under study.
Especially so when considering an adaptive timestep.
In that case, the number of function calls is what determines the number of circuit evaluations and thus also the cost of the algorithm, regardless of the number of accepted or rejected steps.

\section{Results}
\label{sec:results}

\subsection{Noisy variational quantum simulation}
\label{subsec:noise}

\begin{figure*}[t]
  \centering
  \includegraphics{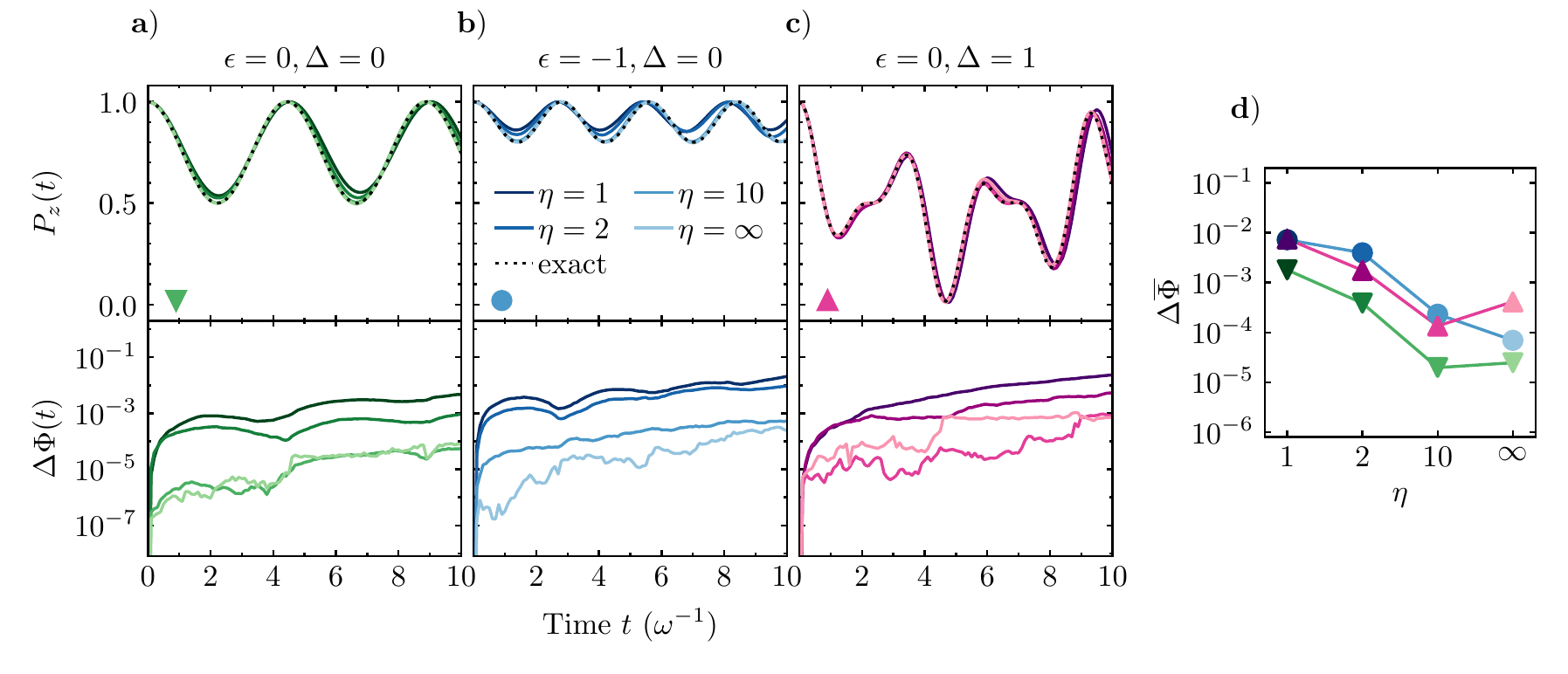}
  \caption{
    Variational simulation results for three qubits ($M=1, n_k^\mathrm{max}=1$), Trotter depth $d=1$, i.e., $N_\theta = 4$ variational parameters, and under varying influence of noise with 8192 shots per circuit evaluation.
    Using the noise model from one of IBM's devices,
    \textit{ibmq\_santiago} v1.3.22
    \cite{ibmqx} (see  \Cref{app:device}, \Cref{tab:santiagoNoise}), $\eta$ denotes the fraction of noise employed.
    That is, $\eta = 1$ indicates results obtained with the full realistic hardware noise together with statistical noise, while $\eta = \infty$ means only statistical and no hardware noise.
    The top row in \textbf{a)-c)} shows the spin-orientation evolving for different system setups (Hamiltonian parameters $\epsilon, \Delta$), while the bottom row shows the respective infidelities of the variational state.
    In panel \textbf{d)}, we plot the mean of the infidelities in \textbf{a)-c)}, $\Delta \overline{\Phi}$, as a function of $\eta$.
    Results indicate that already a reduction of current hardware noise by one order of magnitude yields an accuracy comparable to that obtained with only statistical noise.
  }
  \label{fig:qasmNoise}
\end{figure*}

In the following, we study the spin-boson model with the Hamiltonian in~\Cref{eq:hamiltonian} for various Hamiltonian parameters and system sizes.
In particular, we consider here the resonant case $\omega_k \equiv \omega, g_k \equiv g$ and the regime of ultrastrong coupling (USC), where $g/\omega \in [0.1, 1]$.
Note that, from here on, we will take $H/\omega$ such that all Hamiltonian parameters are expressed in terms of bosonic eigenfrequencies.
We begin with the simplest case of a single bosonic mode with an excitation number cutoff at $\nmax = 1$, resulting in three qubits under the direct mapping.
The coupling strength is fixed at $g / \omega = 0.5$ and we distinguish $(\epsilon, \Delta) \in \{ (0,0), (-1, 0), (0, 1) \}$.
Furthermore, we prepare the initial state in the non-interacting ground state $\ket{01}_\mathrm{b}\ket{0}_\mathrm{s} = \ket{010}$ and monitor its evolution through the orientation of the spin, $P_z = \braket{\sz + 1}/2$.
Note that we use the reverse qubit-ordering notation as conventional in \textsc{Qiskit} \cite{qiskit}.

We aim to investigate how much variational simulations are affected by varying levels of noise.
To this end, we differentiate four regimes; one with statistical (shot) noise only, one with full hardware noise mimicking IBM's \textit{ibmq\_santiago} device \cite{ibmqx}, which belongs to IBM's 5-qubit Falcon processors, as well as two intermediate regimes.
The two intermediate regimes are achieved by mimicking a device through \textsc{Qiskit}'s noise model feature and the possibility to isolate and manipulate specific noise components.
This allows for a detailed study of the influence of current hardware noise.
Particularly, for the intermediate noise regimes, we decrease the average one- and two-qubit gate errors, $e_\mathrm{1qg}$ and $e_\mathrm{2qg}$, respectively, as well as readout errors of the device, $e_\mathrm{read}$, while simultaneously increasing average relaxation and dephasing times, $T_1$ and $T_2$, respectively, by a factor $\eta$,
\bea{C}
    e_\mathrm{1qg} = e_\mathrm{1qg}^\mathrm{dev}/\eta \ , \ e_\mathrm{2qg} = e_\mathrm{2qg}^\mathrm{dev}/\eta \ , \nonumber \\
    e_\mathrm{read} = e_\mathrm{read}^\mathrm{dev}/\eta \ , \\
    T_1 = \eta T_1^\mathrm{dev} \ , \ T_2 = \eta T_2^\mathrm{dev} , \nonumber
\label{eq:noiseReduc}
\eea
To simulate this setup, we employ \textsc{Qiskit}'s shot-based Qasm-simulator with 8192 shots per circuit evaluation, and \textsc{SciPy}'s adaptive Runge-Kutta solver of order 5(4) \cite{2020SciPy-NMeth}.

\Cref{fig:qasmNoise} shows the results of these simulations with $\eta \in \{ 1,2,10,\infty \}$, where $\eta=1$ denotes full hardware noise and statistical noise, whereas $\eta=\infty$ means no hardware noise, i.e., only statistical noise.
We employed complete readout error mitigation \cite{qiskit} via $2^{N_\mathrm{q}}$ calibration circuits where $N_\mathrm{q}$ is the number of qubits.
All evolutions in \Cref{fig:qasmNoise} were obtained with a variational circuit of Trotter depth $d=1$, containing $N_\theta = 4$ variational parameters.
The top row of panels a)-c) displays the evolution of $P_z(t)$, while the respective bottom row gives the infidelities $\Delta \Phi (t) = 1- \left| \braket{\Phi(t)|\Psi(t)} \right|$, where $\Phi(t)$ is the propagated (noisy) variational state at time $t$, while $\Psi(t)$ is the corresponding exact solution obtained by exponentiation of the Hamiltonian matrix.
It is evident that mere statistical noise ($\eta = \infty$) yields high accuracy throughout the entire simulation time, with a final infidelity of $\order(10^{-4})$ to $\order(10^{-3})$.
We note that this is achieved with $\order(10^3)$ integration steps and a total of $\order(10^7)$ shots throughout one simulation.
While this is highly model dependent, we achieve a final accuracy at a fixed number of samples several orders of magnitude better than those estimated in \cite{Barison2021}.

Moreover, although the accuracy decreases with the introduction of hardware noise, the variational algorithm (using the proposed variational ansatz in \Cref{eq:ansatz}) achieves a final infidelity of $\order(10^{-2})$ to $\order(10^{-1})$, even with full hardware noise ($\eta = 1$).
Importantly, despite a deviation of the variational state from the true state trajectory over time, basic physical properties of the system's evolution, such as the oscillation frequency, are reproduced at least qualitatively.

Systematically reducing the noise ($\eta = 2$ and $\eta = 10$) as in \Cref{eq:noiseReduc} gradually increases accuracy.
Remarkably, for $\eta = 10$, the simulation accuracy is comparable to that without hardware noise ($\eta = \infty$).
This becomes even more clear in \Cref{fig:qasmNoise} d), where we plot the mean error $\Delta\overline{\Phi}$ for all $\eta$ and the respective system from a)-c).

\subsection{Scaling up -- simulating larger spin-boson systems}
\label{subsec:5qubits}

\begin{figure*}[t]
  \centering
  \includegraphics{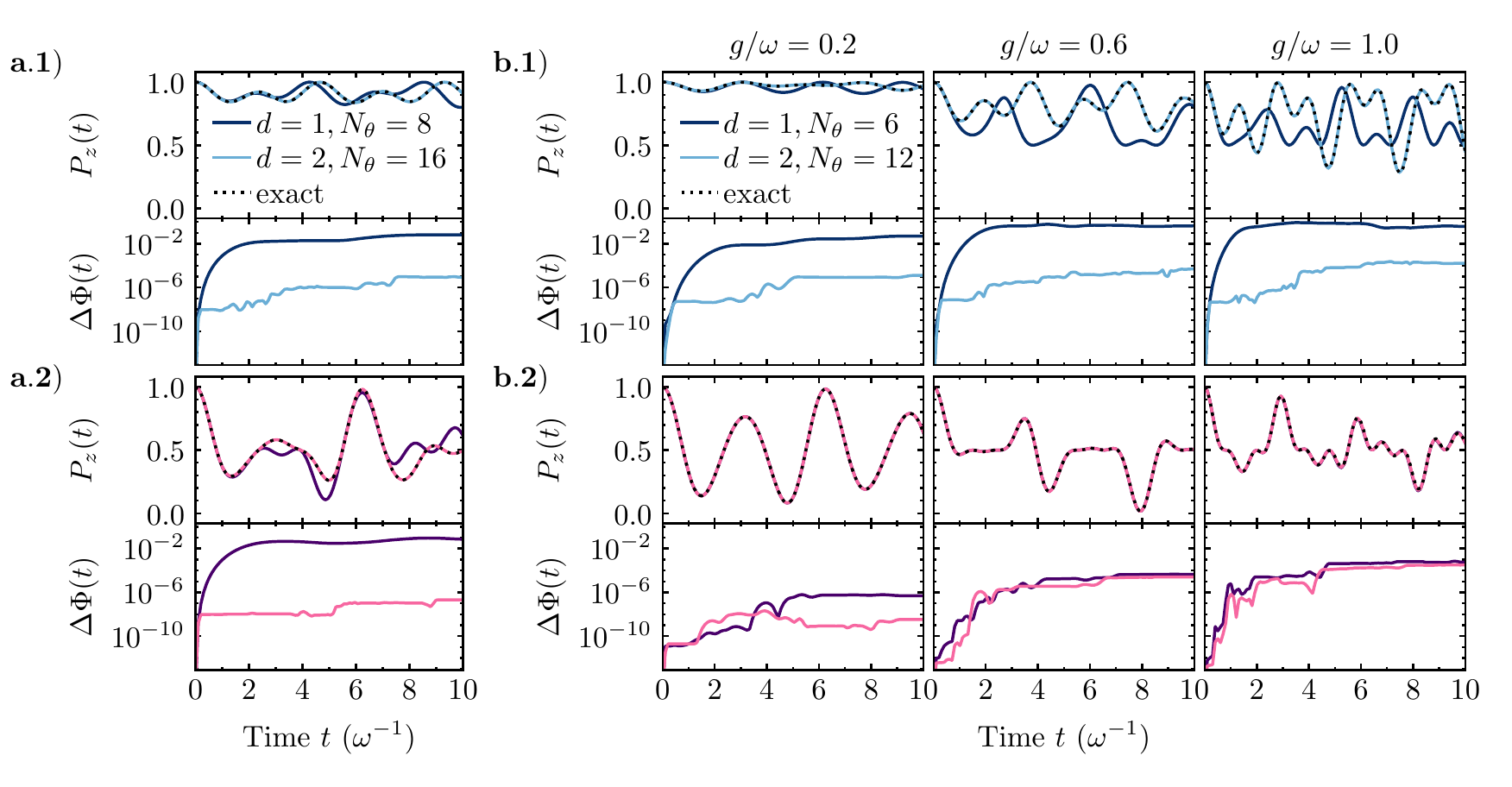}
  \caption{
      \textbf{a)} Variational results for five qubits ($M=1, n_k^\mathrm{max}=3$) and $g/\omega=0.5$, for Trotter depth $d=1, 2$, i.e., $N_\theta = 8, 16$ variational parameters.
      \textbf{b)} Variational results for five qubits ($M=2, n_k^\mathrm{max}=1$) and varying coupling strength $g/\omega=0.2,0.6,1.0$, for Trotter depth $d=1, 2$, i.e., $N_\theta = 6, 12$ variational parameters.
      Top rows \textbf{a.1)}, \textbf{b.1)} correspond to system parameters $(\epsilon, \Delta) = (-1, 0)$, bottom rows \textbf{a.1)}, \textbf{b.1)} the parameters $(\epsilon, \Delta) = (0, 1)$.
  }
  \label{fig:5qubits}
\end{figure*}

Next, we enlarge the system to five qubits, which will enable us to make better scaling predictions for classically intractable systems in \Cref{subsec:advantage}.
Based on the direct qubit-mapping, five qubits may represent two different systems -- a spin coupled to one bosonic mode ($M=1$) with an excitation cutoff at $\nmax = 3$, or a spin coupled to two bosonic modes ($M=2$) with an excitation cutoff at $\nmax = 1$ each.
The time-evolution of these two systems within the same setup as before are displayed in \Cref{fig:5qubits} a) and b), respectively.
Note that, in this and the following subsection, we consider statevector simulations only.
Top rows a.1), b.1) and bottom rows a.2), b.2) represent system parameters $(\epsilon, \Delta) = (-1, 0), (0, 1)$, respectively.

We observe that Trotter depth $d=1$ does not offer enough variational flexibility in all cases anymore and a depth of $d=2$ is necessary to account for the correct dynamics.
This could be anticipated from a simple dimensional analysis of the Hilbert space. 
However, with a total of $N_\theta = 16$ (a) and $N_\theta = 12$ (b) real variational parameters at $d=2$, the dimensionality of the variational state remains well below the exponential size of the full 5-qubit wavefunction, which would require 31 complex or 62 real parameters for full parameterization (two of the $2 \times 2^5$ real parameters may be fixed with norm and global phase).

\subsection{Comparison with Trotter-evolution}
\label{subsec:trotter}

With circuit depth and two-qubit gate-count being the main limiting factors in noisy near-term quantum simulation due to short coherence times, it is worthwhile to compare the variational results from \Cref{fig:qasmNoise,fig:5qubits} to the more resource-intensive Trotter-evolution.
Importantly, in Trotter-evolution, the depth increases with simulation time, while in variational simulation, the ansatz-depth remains constant throughout the simulation.
Henceforth, the question is how the ansatz-depth required by the variational approach scales with system size compared with Trotter-simulation.

To address this question, we use the first-order formula of \Cref{eq:trotter} from \Cref{subsec:productformulas} and aim at finding the minimal Trotter depth $d$ to achieve a final accuracy of $\varepsilon_\mathrm{thresh}$.
For this, we compute the infidelity $\Delta \Phi(t)$ after each Trotter step.
Beginning with a single circuit layer, $d=1$, we append a layer to the circuit every time the infidelity increases above the threshold and repeat the step until $\Delta \Phi(t) < \varepsilon_\mathrm{thresh}$ again.

\begin{figure}[t]
  \centering
  \includegraphics{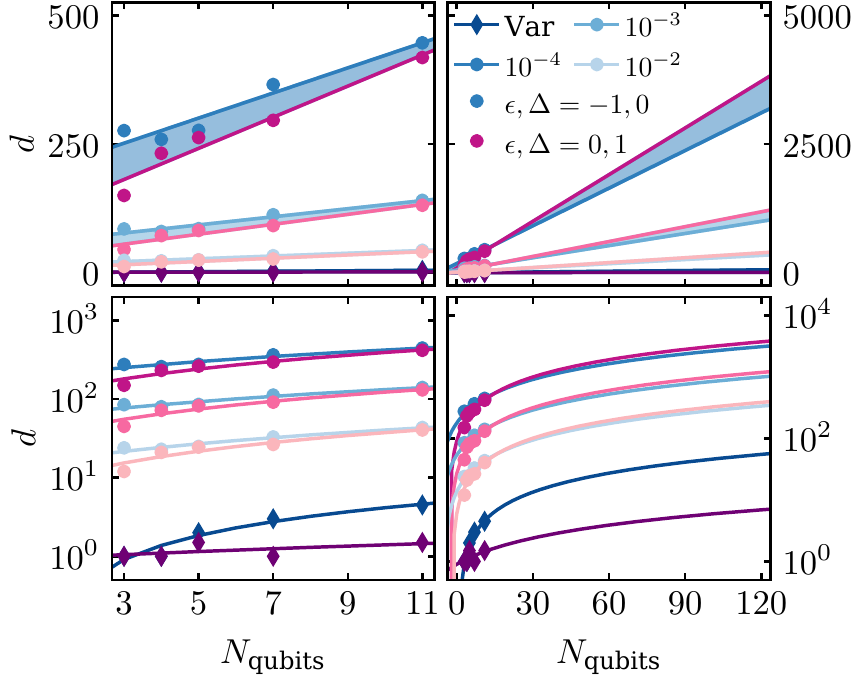}
  \caption{
  Final depth, i.e., number of Trotter steps, required to achieve an accuracy $\Delta \Phi$ below $\varepsilon_\mathrm{thresh}$, comparing Trotter and variational simulation for different system sizes (data points).
  Top and bottom plots show the same data with a linear and log scale, respectively.
  Legend entries denote different values of $\varepsilon_\mathrm{thresh}$ and variational simulation, respectively.
  Variational simulation achieves a final infidelity below $10^{-4}$ throughout all numerical examples and is therefore to be compared to the $\varepsilon_\mathrm{thresh}=10^{-4}$ Trotter curve.
  The scaling of the depth is estimated with a linear fit (see main text), which is used to extrapolate to system sizes of $N_\mathrm{qubits} = 120$, the number of qubits necessary for a simulation comparable to state of the art classical spin-boson simulations.
  }
  \label{fig:trotter}
\end{figure}

The findings shown in \Cref{fig:trotter} emphasize the resource-efficiency of the variational approach when used in combination with a well-chosen ansatz.
Here we plot the final Trotter depth necessary to keep $\Delta \Phi(t) < \varepsilon_\mathrm{thresh}$ throughout a fixed simulation time of $T=10$ with $\varepsilon_\mathrm{thresh} \in \{ 10^{-2}, 10^{-3}, 10^{-4} \}$ and for several system sizes indicated by the number of qubits, $N_\mathrm{q} \in [ 3, \ldots , 11 ]$.
These are compared to the smallest Trotter depth of the variational ansatz that achieves $\Delta \Phi(t) \leq 10^{-4}$ in the variational simulations.
Note that, with a growing number of qubits, the number of distinct spin-boson systems that can be mapped to $N_\mathrm{q}$ increases.
For example, systems with $M=2, \nmax=1$ and $M=1, \nmax=3$ both result in 5 qubits; systems with $M=2, \nmax=4$ and $M=5, \nmax=1$ both result in 11 qubits.
Data points in \Cref{fig:trotter} represent simulation results averaged over all possible systems with the same number of qubits and $\nmax_k \equiv \nmax$.
Detailed numbers may be looked up in \Cref{app:details}, \Cref{tab:depth_eps1,tab:depth_eps0}.
In the variational simulations of up to 11 qubits, $d \leq 4$ Trotter steps sufficed to maintain a target infidelity $\Delta \Phi(t) \leq 10^{-4}$.
On the other hand, using Trotter-evolution, the circuit size grows significantly faster with system size.

To underline the different scaling behaviors, we linearly fit the depths in \Cref{fig:trotter} according to \Cref{eq:trotterScaling}.
Note that $d \propto N_\mathrm{h} \propto N_\mathrm{q}$ since the number of terms in the qubit-mapped Hamiltonian is $N_\mathrm{h} = 7M\nmax +2$.
The fit results are represented by the lines on different scales (linear and log scale in the top and bottom row, respectively, as well as different system size regimes, left and right) and the parameters may be looked up in \Cref{app:details}, \Cref{tab:fit_params}.

Although the depth scales linearly with system size both with Trotter-evolution as well as with the variational approach, it becomes visible from the fitting lines that the scaling coefficients vastly differ in both cases (cf. \Cref{tab:fit_params}).
In fact, despite the obvious savings of the variational method in terms of circuit depth, the extrapolated number of circuit layers for a 120-qubit system is merely around two orders of magnitude smaller than that estimated for Trotter-evolution with a target accuracy of $\varepsilon_\mathrm{thresh}=10^{-4}$.
While this may seem like a large resource saving, it has to be put in relation with additional computational costs associated with the variational scheme, as will become more clear in the following section.

\section{Scaling estimates for quantum advantage: variational approach vs. Trotter}
\label{subsec:advantage}

The world's largest supercomputers can store in the order of $10^{12}$ bits of information.
This corresponds, for instance, to the size of the Hilbert space of a 12-mode spin-boson systems with 10 degrees of freedom per mode ($M=12, \nmax=9$).
For the simulation of such a system as described in the previous sections, we would need to control $N_q = 121$ qubits ($122$ with an ancilla in variational simulation).
In the following, we want to estimate the computational effort necessary to simulate such a system with both the variational and the Trotter-based approach, making use of the scaling laws presented in \Cref{subsec:resources} and the fits reported in \Cref{fig:trotter}.
Throughout this section, we will consider a maximal target error of $\varepsilon \leq 10^{-4}$ during the entire simulation.

From the results in \Cref{fig:trotter}, we estimate that the propagation of such a system using the first-order product formula (\Cref{eq:trotter}) will require a circuit depth $d \approx 3400$ in order to achieve the desired accuracy. 
This value is obtained by taking the average prediction for both Hamiltonian regimes in \Cref{fig:trotter}, leading to $N_\mathrm{cx} \approx 10^7$ two-qubit gates.
Importantly, this is the only computational cost associated with Trotter-based simulation in order to reach the fixed final time of $T=10$, as no classical data processing is required.

On the other hand, in variational simulations the cost is split into three main components: the one associated to the circuit length (gate counts), the number of circuit evaluations to compute the different matrix elements, and the classical data processing to obtain the parameter update.
The same extrapolation based on \Cref{fig:trotter} predicts a variational form with a depth $d\approx 31$ to simulate $N_q=122$ qubits.
In this case the number of 2-qubit gates amounts to $N_\mathrm{cx} \approx 10^5$.
With a final time of $T=10$ and a desired accuracy of $\varepsilon \leq 10^{-4}$, the total number of timesteps $N_t = \varepsilon / \varepsilon_\mathrm{local}$ (cf. \Cref{eq:timesteps}), needed to integrate the \gls{eom}, amounts to $N_t \approx 100$.
Note that in the numerical studies presented above, we found good agreement with this scaling even with an adaptive timestep when carefully choosing absolute and relative error tolerances for acceptance criteria.
We report exact numbers of function calls for statevector as well as noisy simulations in \Cref{app:details}, \Cref{tab:funCalls}.
The total number of circuit evaluations for the variational case becomes 
\begin{align}
    N_\mathrm{circ}^{\mathrm{tot}} &= N_t  N_\mathrm{circ} = \order \bigg( \frac{N_t}{\varepsilon^2} \bigl( N_{\diff \theta}^2+ N_\mathrm{h} N_{\diff \theta} \bigr) \bigg) \\
    &\approx 10^{18} \ . \nonumber
\end{align}

Now, assuming a two-qubit gate length of \SI{100}{\nano\second}, executing the Trotter-circuit takes approximately $1 \, \mathrm{second}$.
A single evaluation of the variational circuit, on the other hand, would last roughly $0.01 \, \mathrm{seconds}$.
We neglect the additional time required to measure and reset qubits after each circuit evaluation and the speed-up from possibly parallelizing circuit evaluations in the variational approach, since these two effect counter each other.
Under these assumptions, a variational simulation will take approximately $\SI{0.01}{\second} \times 10^{18} = \SI{e15}{\second} \approx 3 \times \SI{e7}{\year}$.

This example illustrates that, although the variational procedure allows for shallower circuits and hence opens avenues for performing simulations of small systems on near-term quantum computers, it is unlikely that it will lead to quantum advantage for the simulation of spin-boson models.
In fact, the number of circuit evaluations quickly becomes prohibitive in this case.
This issue was also raised by Barison and coworkers~\cite{Barison2021} who proposed a variational algorithm which reduces the scaling of $N_{\mathrm{circ}}$ from quadratic to linear in the number of parameters.
Although this step goes in the right direction, it is by itself not enough to make the variational approach feasible for the applications described here.
It is worth mentioning that the scaling could be further reduced by improving the sampling procedure.
However, for an optimal number of shots, $N_\mathrm{shots} = \mathcal{O}(\log(1/\varepsilon))$, and a linear scaling of $N_{\mathrm{circ}}$ with the number of parameters, the number of circuit evaluation would still amount to $N_\mathrm{circ}^{\mathrm{tot}} \approx 10^{10}$, taking roughly $\SI{e7}{\second} \approx \SI{0.3}{\year}$.
\textcolor{black}{At the same time, the scaling of product formulas is sub-optimal and novel algorithms exhibiting reduced complexity have been proposed recently.
Most notably, qubitization~\cite{Low2019hamiltonian} achieves a gate complexity linear and additive in time, $\order (t + \log(1/\epsilon))$, which is provably optimal.
A rough estimate based on the asymptotic bounds presented in Ref.~\citenum{Low2019hamiltonian} suggests that qubitization requires a two-qubit gate count of $N_\mathrm{cx} = \order (10^5)$ for the above example, taking $\order (\SI{0.01}{\second})$ to run, and additional $\order (10)$ ancilla qubits to implement the needed oracles.
Despite its potential, the implementation of this algorithm poses important challenges to near-term quantum computing, which cannot be addressed in this work.}

\section{Discussion and Conclusions}
\label{sec:conclusion}

In this work, we investigated the performance of a time-evolution \gls{vqa} by simulating the quantum dynamics of a spin-boson Hamiltonian, a model which is widely used  to describe the embedding of a two-level system in a thermal bath.
Aside from assessing the \gls{vqa}'s numerical stability, the purpose of our investigation is to provide scaling estimates and predictions, particularly in comparison to conventional Trotter-evolution.
In particular, we analyzed the performance of these time-evolution algorithms in the regimes of near-term and fault-tolerant quantum computing.

To this end, we studied the dynamics of several spin-boson systems, varying in size as well as in the Hamiltonian parameter space (i.e., the Hamiltonian coefficients).
Furthermore, we introduced hardware noise into the variational simulations by using the noise model of one of IBM's quantum computers, which provided a clear upper bound for the level of noise tolerated by the algorithm.
Throughout all simulations, the physically motivated variational ansatz, which we constructed based on the system Hamiltonian, offered a great deal of flexibility and correctly captured various system's dynamics without the need of further tuning the variational quantum circuits.
Moreover, it exhibits linear scaling of both the number of variational parameters and circuit depth.

Concerning the scaling of the two considered methods for time-evolution, namely the variational and the Trotter-based approach, we presented approximate scaling laws for the classical and the quantum computational resources required by both methods.
We further performed a series of simulations for system sizes in the range $N_\mathrm{q} \in \{3, \ldots, 11 \}$ to determine the required circuit depths for a fixed target error, and extrapolated these values to larger numbers of qubits using appropriate fitting models.
Based on these extrapolations, we could estimate the computational cost of both methods for simulating system sizes, which are barely accessible with cutting-edge classical algorithms.

From this analysis, we can conclude that the variational approach in the current implementation is an efficient and reliable approach in the case of relatively small setups, especially in the context of current hardware limitations.
However, the costs associated with the number of circuit evaluations will quickly become unaffordable, hampering its applicability to large setups.
Note that this occurs despite the fact that the number of resources (variational parameters and gate count) only increases linearly with the system size.
Although Trotter-evolution has a two-qubit gate count two orders of magnitude larger than the variational method, it does not suffer from the same prohibitive scaling of required measurements.

In conclusion, the variational algorithm might be a useful tool for demonstrations of small system's dynamics on noisy near-term quantum devices.
But it remains an open issue whether or not, and if so, under which circumstances, it would potentially become a valid alternative to the Trotter-based approach for quantum dynamics simulations of systems with many degrees of freedom.
At least in the simulation of the spin-boson model, the Trotter-based algorithm remains superior to the variational approach for treating system sizes currently intractable with classical computers.

\subsection*{Acknowledgements}
The authors thank Christa Zoufal, Francesco Tacchino, Irene Burghardt, and Rocco Martinazzo for their help and inspiring discussions.\\
The authors acknowledge financial support from the Swiss National Science Foundation (SNF) through the grant No. 200021-179312.\\
IBM, the IBM logo, and ibm.com are trademarks of International Business Machines Corp., registered in many jurisdictions worldwide. Other product and service names might be trademarks of IBM or other companies. The current list of IBM trademarks is available at \url{https://www.ibm.com/legal/copytrade}.\\



\bibliography{refs.bib}

\pagebreak

\appendix

\begin{figure}[t]
  \includegraphics{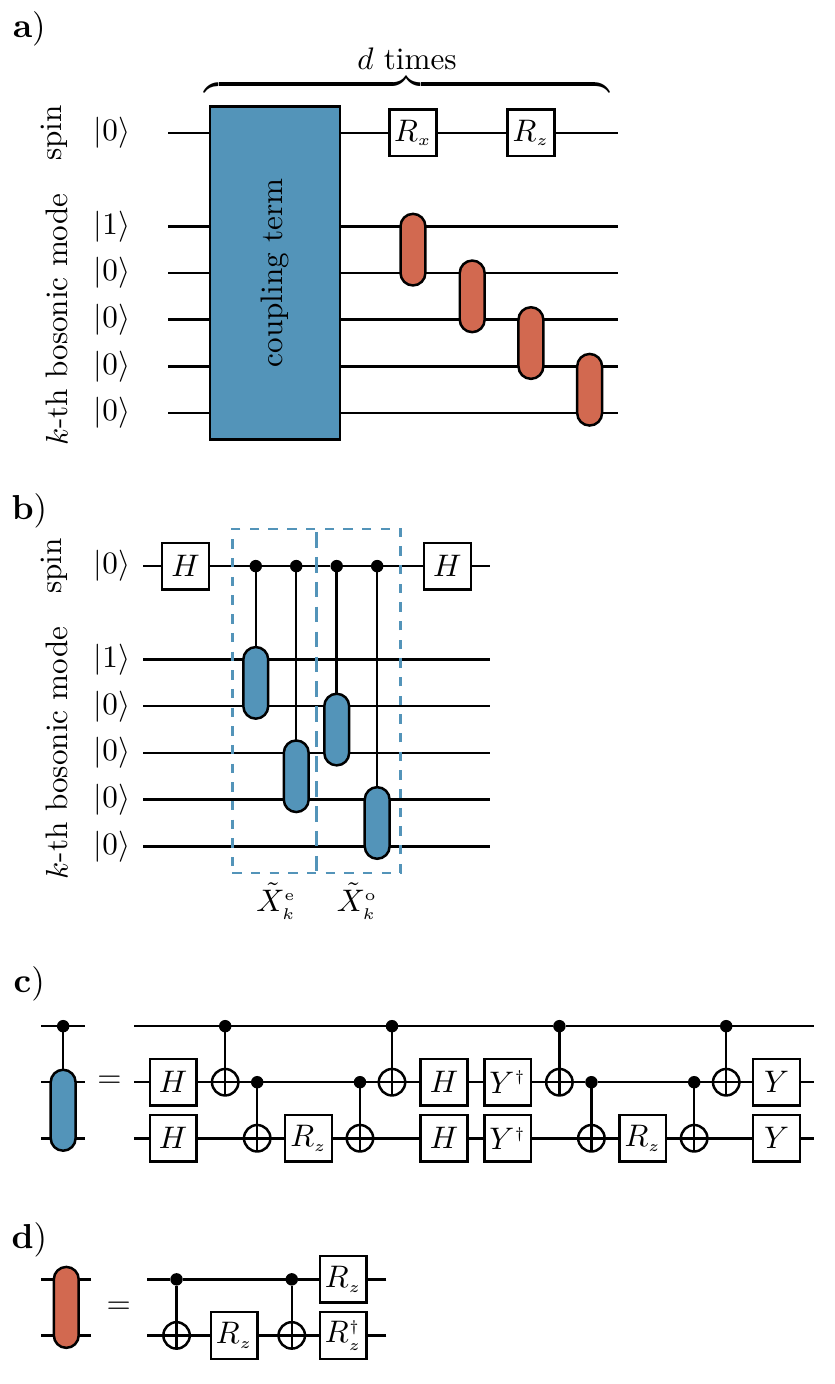}
  \caption{
    \textbf{a)} Quantum circuit representation of the variational ansatz $U_\mathrm{H}$ in \Cref{eq:ansatz}, showing the spin and one bosonic qubit register.
    The blue box represents the coupling term in $U_\mathrm{H}$, while the red two-qubit gates represent bosonic self-interaction terms.
    \textbf{b)} Decomposition of the coupling gate from \textbf{a)}, with each cascade of three-qubit gates representing a term in $\tilde{X}_k^\mathrm{e,o}$.
    \textbf{d)} Final decomposition of the three-qubit gate into one- and two-qubit gates, coupling the spin and the respective bosonic mode.
    \textbf{c)} Decomposition of the bosonic self-interaction gate.
    Variational parameters enter through rotational gates $R_z(\theta)$.
    The Hadamard gate $H$ and $Y^\dagger = R_x(\pi/2)$ rotate qubits from $\sigma^z$-- into the $\sigma^x$-- and $\sigma^y$--basis, respectively.
  }
  \label{fig:varcirc}
\end{figure}


\section{The variational quantum circuit}
\label{app:varcirc}

In this section, we detail the construction of the quantum circuit from the qubit-mapped ansatz $U_\mathrm{H}$ in \Cref{subsec:varform}.
A detailed representation for one bosonic mode is shown in \Cref{fig:varcirc}, whereby the circuit for $M$ bosonic modes is obtained by appending resepective bosonic qubit registers that are coupled to the qubit representing the spin via respective coupling gates (blue box in \Cref{fig:varcirc} a) and which have their own self-interaction gates (red two-qubit gates in \Cref{fig:varcirc} a).

For all simulations in the main-text, we initialize each bosonic register in its non-interacting ground state $\ket{\tilde{0}}_k = \ket{0 \ldots 01}$, obtained with a bit-flip on the first $k$-mode qubit, $\ket{1} = X \ket{0} = \sx \ket{0}$.

Variational parameters enter through rotational gates as $R_z(\theta) = \exp(-i \theta \sigma^z / 2)$.
Moreover, all parameters are $n_k$-dependent, $\theta = \theta_{n_k}$, such that each bosonic self-interaction and coupling gate (red two- and blue three-qubit gates) is parameterized individually.
Within each of these gates, however, for instance within the gate in c), all three $R_z$-gates have the same parameter.
While these parameters could be made independent as well if more flexibility is required of the ansatz, we found no advantage in doing so.

\section{Hardware specifications}
\label{app:device}

\begin{table}[t!]
\centering
\begin{ruledtabular}
    \begin{tabular}{L{0.9cm} R{0.95cm} R{1cm} R{1.2cm} R{1.2cm} R{1.4cm}  R{1.0cm}}
        Qubit & $T_1$ & $T_2$ & $e_\mathrm{ro}$ & $e_\mathrm{1qg}$ & $e_\mathrm{2qg}$ & len \\
        & $(\SI{}{\micro\second})$ & $ (\SI{}{\micro\second})$ & $(\times 10^{-2})$ & $(\times 10^{-4})$ & $(\times 10^{-3})$ & ($\SI{}{\nano\second}$) \\
        \midrule
        Q0 & 150.48 & 284.71 & 2.31 & 2.68 & $[0,1]$ 6.25 & 526.22 \\
        Q1 & 163.37 & 104.22 & 1.14 & 1.71 & $[1,0]$ 6.25 $[1,2]$ 6.01 & 561.78 355.56\\
        Q2 & 144.89 & 97.87 & 1.47 & 2.25 & $[2,1]$ 6.01 $[2,3]$ 6.15 & 320.00 376.89 \\
        Q3 & 230.80 & 97.37 & 0.52 & 1.52 & $[3,2]$ 6.15 $[3,4]$ 5.69 & 412.44 376.89 \\
        Q4 & 47.22 & 103.46 & 2.16 & 3.81 & $[4,3]$ 5.69 & 341.33 \\
        \midrule
        mean & 122.55 & 149.53 & 1.63 & 2.09 & 7.78 & 536.89 \\
    \end{tabular}
\end{ruledtabular}
\caption{Decoherence time, readout as well as one- and two-qubit gate errors of \textit{ibmq\_santiago}, accessed on \textit{June 14, 2021}.
2-qubit gate errors and lengths are listed together with the respective qubit pair, e.g., [0, 1] for the 2-qubit gate between qubits Q0 and Q1.
The mean values were used for studying different levels of hardware noise in \Cref{subsec:noise}.}
\label{tab:santiagoNoise}
\end{table}

\Cref{tab:santiagoNoise} lists the most relevant device specifications of all 5 qubits of the device used in \Cref{subsec:noise}, \textit{ibmq\_santiago, v1.3.22}.
It is important to note that this is just a snapshot of the device's noise and that, in reality, these quantities may vary.

\newpage

\section{Details of numerical experiments}
\label{app:details}

Here, we report relevant details of the simulations described the main text.
\Cref{tab:depth_eps0,tab:depth_eps1} list the Trotter depths of the circuits shown in \Cref{fig:trotter}, and \Cref{tab:fit_params} details the corresponding fit parameters.
Lastly, \Cref{tab:funCalls} lists the number of function calls by the adaptive-step solver in \Cref{fig:qasmNoise}.

\begin{table}[h!]
\centering
\begin{ruledtabular}
    \begin{tabular}{R{0.5cm} C{1.5cm} R{1.0cm} R{1.0cm} R{1.0cm} R{0.8cm}}
        $N_\mathrm{q}$ & $(M, \nmax)$ & $10^{-2}$ & $10^{-3}$ & $10^{-4}$ & Var \\
        \midrule
         3 & $(1,1)$ & 24 & 85 & 276 & 1 \\
         4 & $(1,2)$ & 23 & 80 & 259 & 1 \\
         5 & $(1,3)$ & 23 & 73 & 230 & 2 \\
           & $(2,1)$ & 27 & 98 & 323 & 2 \\
         7 & $(2,2)$ & 34 & 111 & 356 & 3 \\
           & $(3,1)$ & 32 & 114 & 375 & 3 \\
        11 & $(2,4)$ & 39 & 124 & 389 & 4 \\
           & $(5,1)$ & 48 & 157 & 504 & 5 \\
    \end{tabular}
\end{ruledtabular}
\caption{Trotter depth $d$, necessary to simulate spin-boson systems of different size ($M, n^\mathrm{max}$) and $\epsilon = -1, \Delta = 0$.
These numbers are plotted in \Cref{fig:trotter} in the main text.
Columns correspond to simulations using Trotter-evolution with three different values of final accuracy $\varepsilon_\mathrm{thresh} \in \{ 10^{-2}, 10^{-3}, 10^{-4} \}$, and using variational simulation, respectively.}
\label{tab:depth_eps1}
\end{table}

\begin{table}[h!]
\centering
\begin{ruledtabular}
    \begin{tabular}{R{0.5cm} C{1.5cm} R{1.0cm} R{1.0cm} R{1.0cm} R{0.8cm}}
        $N_\mathrm{q}$ & $(M, \nmax)$ & $10^{-2}$ & $10^{-3}$ & $10^{-4}$ & Var \\
        \midrule
         3 & $(1,1)$ & 12 & 45 & 150 & 1 \\
         4 & $(1,2)$ & 21 & 72 & 232 & 1 \\
         5 & $(1,3)$ & 31 & 98 & 311 & 2 \\
           & $(2,1)$ & 18 & 66 & 214 & 1 \\
         7 & $(2,2)$ & 30 & 102 & 329 & 1 \\
           & $(3,1)$ & 23 & 81 & 263 & 1 \\
        11 & $(2,4)$ & 50 & 157 & 496 & 2 \\
           & $(5,1)$ & 31 & 105 & 340 & 1 \\
    \end{tabular}
\end{ruledtabular}
\caption{Same as \Cref{tab:depth_eps1} but for $\epsilon = 0, \Delta = 1$.}
\label{tab:depth_eps0}
\end{table}

\begin{table}[h!]
\centering
\begin{ruledtabular}
    \begin{tabular}{R{1cm} L{0.5cm} R{1.1cm} R{1.1cm} R{1.1cm} R{0.9cm}}
        $\epsilon, \Delta$ & fit params & $10^{-2}$ & $10^{-3}$ & $10^{-4}$ & Var \\
        \midrule
         $-1,0$ & $p_1$ $p_0$ residual & $2.69$ $13.58$ $11.89$ & $7.86$ $53.53$ $156.54$ & $24.39$ $178.36$ $1721.29$ & $0.46$ $-0.48$ $0.24$ \\
         $0,1$ & $p_1$ $p_0$ residual & $3.16$ $5.93$ $27.64$ & $9.64$ $26.46$ $218.54$ & $30.24$ $90.28$ $1893.54$ & $0.05$ $0.90$ $0.20$ \\
    \end{tabular}
\end{ruledtabular}
\caption{Results of a linear fit $f(x) = p_1 x + p_0$ to the data points in \Cref{fig:trotter}.}
\label{tab:fit_params}
\end{table}

\begin{table}[h!]
\centering
\begin{ruledtabular}
    \begin{tabular}{R{1cm} C{1.cm} R{1.0cm} R{1.0cm} R{1.0cm} R{0.8cm}}
        $\epsilon, \Delta$ & SV & $\eta = \infty$ & $\eta = 10$ & $\eta = 2$ & $\eta = 1$ \\
        \midrule
         $0,0$ & 182 & 5282 & 2840 & 710 & 506 \\
         $-1,0$ & 428 & 3554 & 1670 & 578 & 482 \\
         $0,1$ & 230 & 9518 & 1892 & 890 & 596 \\
    \end{tabular}
\end{ruledtabular}
\caption{We report the number of function evaluations performed by \textsc{SciPy}'s RK45 solver with adaptive timestep \cite{2020SciPy-NMeth} for the $M=1, n^\mathrm{max} = 1$ system in \Cref{fig:qasmNoise}.
For reference, we include the numbers for statevector simulations of the same systems (SV).
Since these numbers are highly dependent on numerical tolerances, we report here also the absolute and relative error tolerance for choosing the step size, $\delta_\mathrm{a} = 10^{-6}$ and $\delta_\mathrm{r} = 10^{-3}$, respectively for statevector, and $\delta_\mathrm{a} = 10^{-3}, \delta_\mathrm{r} = 10^{-3}$ for noisy simulations.
Furthermore, the singular value cutoff for matrix inversion (necessary in solving \Cref{eq:McLachlanEOM}) was fixed at $\delta_\mathrm{cond} = 10^{-6}$ and $\delta_\mathrm{cond} = 10^{-3}$ for statevector and noisy simulations, respectively.}
\label{tab:funCalls}
\end{table}

\newpage

\end{document}